\documentclass[prl,twocolumn,superscriptaddress]{revtex4}

\usepackage{comment}
\usepackage{amsmath,amssymb}
\usepackage{verbatim}
\usepackage{graphicx}
\usepackage{mathrsfs,yfonts}  
\usepackage[
      colorlinks=true,
      linkcolor=blue,
      urlcolor=magenta,
      filecolor=green,
      citecolor=red,
      pdfstartview=FitV,
      pdftitle={},
        pdfauthor={A.A. Abolhasani, M.M. Sheikh-Jabbari and M.H. BazrAfshan},
        pdfsubject={},
        pdfkeywords={},
        pdfpagemode=None,
        bookmarksopen=true
      ]{hyperref}
\usepackage{color}

\DeclareFontFamily{OT1}{rsfs}{}
\DeclareFontShape{OT1}{rsfs}{m}{n}{ <-7> rsfs5 <7-10> rsfs7 <10->rsfs10}{} 
\DeclareMathAlphabet{\mycal}{OT1}{rsfs}{m}{n}

\newcommand{\bel}[1]{ \begin{equation}\label{#1} }
\newcommand{\p}{\partial}

\newcommand{\gsim}{\lower.7ex\hbox{$\;\stackrel{\textstyle>}{\sim}\;$}}
\newcommand{\lsim}{\lower.7ex\hbox{$\;\stackrel{\textstyle<}{\sim}\;$}}

\def\mpl{M_{\rm Pl}}
\newcommand{\be}{\begin{equation}}
\newcommand{\ee}{\end{equation}}
\newcommand{\bea}{\begin{eqnarray}}
\newcommand{\eea}{\end{eqnarray}}

\newcommand{\expect}[1]{\left\langle #1 \right\rangle}

\newcommand{\bsb}{\boldsymbol}
\newcommand{\nn}{\nonumber}
\newcommand{\vphi}{\varphi}

\def\p{{\bsb p}}
\def\k{{\bsb k}}

\def\K{{\bsb k}}

\newcommand{\bcm}{}

\newcommand{\bi}{\begin{itemize}}
\newcommand{\ei}{\end{itemize}}
\newcommand{\bt}{\begin{tabular}}
\newcommand{\et}{\end{tabular}}
\newcommand{\bc}{\begin{center}}
\newcommand{\ec}{\end{center}}

\def\one{{\hbox{ 1\kern-.8mm l}}}
\newcommand{\Dslash}{\not{\hbox{\kern-4pt $D$}}}
\newcommand{\pdslash}{\not{\hbox{\kern-2pt $\partial$}}}
\newcommand{\ba}{\begin{array}}
\newcommand{\ea}{\end{array}}

\def\bbox{{\,\lower0.9pt\vbox{\hrule \hbox{\vrule height 0.2 cm
\hskip 0.2 cm \vrule height 0.2 cm}\hrule}\,}}
\newcommand{\dsl}{\pa \kern-0.5em /}

\definecolor{summersky}{cmyk}{0.71,0.33,0,0.14}
\definecolor{flamingo}{cmyk}{0,0.51,0.71,0.14}

\newcommand\snote[1]{\textcolor{blue}{\bf [Shahin:\,#1]}}

\begin{document}

\newcommand{\mytitle}{{Resonant Reconciliation of Convex Models and the Planck}}

\title{\mytitle}

\author{A.A. Abolhasani}
\email{abolhasani@ipm.ir}
\affiliation{Department of Physics, Sharif University of Technology, Tehran, Iran}

\author{M.M. Sheikh-Jabbari}
\email{jabbari@theory.ipm.ac.ir}
\affiliation{School of physics, Inst. for research in fundamental sciences (IPM), P.O.Box 19395-5531, Tehran, Iran}


\date{\today}

\preprint{IPM/P-2019/nnn}

\begin{abstract}

We consider single field chaotic inflationary models plus a cosine modulation term, as in axion monodromy models, and augment it by a light scalar field with similar cosine coupling. We show the power spectrum of curvature perturbations of this model is dominated by the one-loop contribution to inflaton two-point function which is enhanced due to resonant interactions. This allows to disentangle the scale of scalar and tensor perturbations and hence to suppress the ratio of tensor-to-scalar power spectra and alters the expression of scalar spectral tilt from the simple chaotic models, thus opening the way to reconcile chaotic models with convex potential and the  Planck data. As in monodromy inflation models, we also have a cosine modulation in spectral tilt. We mention that contribution of resonance effects on non-Gaussianty is small and it remains within the current bounds. Resonant production of light particles toward the end of inflation may set the stage for a  successful reheating model.

\end{abstract}

\pacs{xxxxx}

\maketitle

\setcounter{footnote}{0}

CMB temperature anisotropy observations and other sets of cosmological data \cite{Planck-2018} are supporting inflation as the leading paradigm for the early universe. In simple models of inflation power spectrum of primordial gravity-waves, the tensor modes, sets the scale of inflation; i.e. the value of Hubble parameter during inflation $H$. Nonetheless, the tensor modes have not been observed yet and there is an upper bound  usually reported through tensor-to-scalar ratio $r\lesssim 0.1$. The other parameters reported up to order $10^4$ precision in the Planck data are the spectral tilt $n_s$, or first and second tilts of the tilt \cite{Planck-2018}. The $r-n_s$ plot  is usually used to discriminate models of inflation. 
The remarkable precision of the cosmic data has already ruled out simple models of slow-roll inflation. In particular, models of single field inflation with convex potential has been disfavored \cite{Planck-2018}.

One of the features of single field slow-roll inflation models is that $H$ does not explicitly appear in expressions for $r, n_s$ and these two are only functions of the slow-roll parameters $\epsilon, \eta$ which are given in terms of derivatives of the inflaton potential. This, together with requiring enough number of e-folds of inflation $N_e$ (usually $N_e\gtrsim 60$) is the basic fact allowing for ruling out such models. 

Axion monodromy inflation is among models which has theoretical support and is consistent with the current data \cite{monodromy-SW, axion-monodromy-GW, axion-monodromy-CMB-features, monodromy-and-Planck, monodromy-MSWW, Ngy-axion-monodromy, Planck-2015-inflation, omega-bound}. This model has a (concave) power-law potential, usually a potential linear in inflaton, plus an oscillatory modulation term. The oscillatory part induces cosine-log undulations in the spectral tilt $n_s$ and the non-Gaussianity parameter $f_{NL}$ which could be used to restrict these models \cite{Planck-2015-NG, omega-bound}.  

In this paper we employ ideas of monodromy inflation, namely a single field inflation model plus a subdominant oscillatory term and augment the model by the addition of a light scalar field which couples to inflaton through an ``oscillatory interaction term'' involving cosine of inflaton. As we will argue this can have profound effects on the cosmic perturbation theory of the model, while not backreacting on the inflationary background trajectory. We show that in our model the power spectrum of scalar perturbations is not a function of $H$ and its spectral tilt differs from the usual single field expressions, while the power spectrum of tensor perturbations $P_T$ has the usual expression proportional to $H^2$. This enables us to suppress tensor-to-scalar ratio $r$ and hence relax the bounds on the basic model of inflation, including the models with convex potentials. In the rest of this Letter we show how the model works.

\vspace{-6mm}

\section{The basic setup}\label{sec:2}
\vspace{-4mm}
 
Chaotic inflation models with cosine modulation, which contains family of axion monodromy inflation models, is driven by a periodically modulated polynomial potential for the axion field $\phi$ which plays the role of  inflaton and is governed by the Lagrangian \cite{monodromy-SW,axion-monodromy-GW}
\be
\hspace*{-3mm}{\cal L}_\phi=-\frac12(\partial\phi)^2-V_0(\phi),\ V_0(\phi)=\Lambda^4\left( (\frac{\phi}{ f})^p+b\cos\frac{\phi}{f}\right).
\ee
The model has two parameters of dimension of mass, $\Lambda, f$ and two dimensionless parameters $b$ and $p$. Inflation is driven by $\phi^p$ part and the oscillatory $\cos(\phi/f)$ part induces modulations on the observable signals like on the spectral tilt or non-Gaussianity. The $p=1$ in particular and $p<2$ cases in general, are the usual axion monodromy model which has strong theoretical motivations from string theory \cite{monodromy-SW,axion-monodromy-GW}.
In this work we keep an open eye on $p$, and allow $p>2$ values and keep it as a parameter to be fixed by observational requirements.

The background slow-roll dynamics is governed by 
\be\label{monodromy-slow-roll}
\dot\phi\simeq -\frac{p}{3}\frac{\Lambda^4}{f H}(\frac{\phi}{ f})^{p-1},\qquad \rho_{bg}=3\mpl^2H^2=\Lambda^4(\frac{\phi}{f})^p.
\ee
Validity of the slow-roll approximation is governed by smallness of the slow-roll parameters
\be\label{slow-roll-parameters}\begin{split}
\epsilon&=\frac12\left(\frac{\mpl V_0'}{V_0}\right)^2\simeq \frac{p^2}{2}\big(\frac{\mpl}{\phi}\big)^2,\cr 
\eta &=\frac{\mpl^2V_0''}{V_0}\simeq p(p-1)\big(\frac{\mpl}{\phi}\big)^2- b \frac{\mpl^2 f^{p-2}}{\phi^p}\cos\frac{\phi}{f}.
\end{split}\ee
Number of e-folds of the model is given by 
\be
\hspace*{-3mm}N_e=\int_{\phi_f}^{\phi_i} \frac{1}{\sqrt{2\epsilon}} \frac{d\phi}{\mpl},\qquad \big(\frac{\mpl}{\phi}\big)^2\simeq \frac{1}{2pN_e(\phi)}.
\ee
As the inflaton rolls down the $\phi^p$ potential, the oscillatory part oscillates with frequency $\omega=\dot\phi/f$, and 
\be\label{H/w}
\alpha\equiv \frac{\omega}{H}=\sqrt{2\epsilon}\frac{\mpl}{f}.
\ee
During slow-roll evolution $\omega$ and $H$ are almost constants. Observationally,  modulations around the linear potential, through the cosine term in $\eta$ \eqref{slow-roll-parameters}, induces undulations on the CMB spectrum and bispectrum of the cosine-log form  \cite{axion-monodromy-GW, axion-monodromy-CMB-features, monodromy-and-Planck, Planck-2018}, which yields
$\log_{10}(\omega/H)\sim 1.5-2.1$ \cite{omega-bound}.

\textbf{\textit{The Resonating Module.}}
We now augment our model with the addition of a  light scalar field $\chi$ which couples to the inflaton $\phi$,
 \be 
{\cal L}_{\chi} = -\frac12(\partial\chi)^2- \frac12 m^2_{\chi}(\phi) \chi^2,
 \ee
where the modulated mass term can be the most general coupling which respects the Lorentz and discrete shift symmetries of the $\phi$ field, $\phi\to \phi+2\pi f n$ $n\in \mathbb{Z}$, as these symmetries are protected perturbatively \cite{monodromy-SW,axion-monodromy-GW}. In the language of effective field theories, this means that \footnote{Note that such terms in the effective potential are coming from tree level diagrams where we can suppress the exchanged virtual particles. The expansion becomes more exact if $|p|\ll M$ or $|p|\gg M$, where $|p|$ is the momentum of the exchanged particle and $M$ is its mass. In the former (latter) case $\mathcal{G}$ is a polynomial in positive (negative) powers of $(\partial_{\mu} \phi)^2/f^4$.}
\be 
\label{chi:Mass}
m^2_{\chi}(\phi) = \mu^2+\mathcal{G} f^2 \cos \frac{\phi}{f} ,\ \mathcal{G}\equiv \mathcal{G}\left((\partial_{\mu} \phi)^2/f^4\right)
\ee
In particular,
\be    
\gamma \equiv \dfrac{\dot{\phi}}{f^2}= \frac{\omega}{f},\quad \mathcal{G}=\mathcal{G}(\gamma),
\ee
$\gamma=\omega/f$ is almost a constant during slow-roll period. In our analysis below we keep an open eye on ${\cal G}$ as a function of $\gamma$. As we will see the derivatives $d^n\ln{\cal G}/d(\ln\gamma)^n$ for $n=1,2,3$ are fixed by spectral tilt $n_s$ and first and second tilts of the tilt.

The modulated mass term has essentially two different physical effects: (1) a copious, resonant production of relativistic $\chi$ particles during inflation; (2) introduction of a modulate $\phi-\chi$ coupling, which affects cosmic perturbation theory and power spectra.
In what follows we explore  these two effects and their physical implications.

\textbf{\textit{Creating Relativistic Particles.}}
The equation of motion for mode $k$ of $\chi(t)$ field is
\be\label{chi-eom}
\ddot\chi_k+3H\dot\chi_k+(k^2+m_\chi^2(\phi))\chi_k=0.
\ee
We take the $\chi$ field to be light in the sense that the oscillatory part generically dominates over the $\mu^2$ part, i.e. $\mu^2\ll {\cal G}f^2$. 
Due to the field-dependence of the mass, as the inflaton field $\phi$ rolls towards the minimum of its potential the matter field $\chi$ experiences a mass modulation with the same frequency as the background inflaton $\omega$.
This leads to resonant production of $\chi$-particles and we should hence make sure that this does not affect the background inflationary  trajectory.

For the case of our interest $\alpha\gg 1$, the friction term $3H\dot\chi$ can be safely neglected and  \eqref{chi-eom} reduces to 
\be
\chi ''_{\k}+\left(A_{\k}+2 q \,\cos 2z \right)\chi_{\k}=0,
\ee
where prime denote derivative w.r.t. $z=\omega t/2$ and
\be
A_{\k}=4 \dfrac{k^2}{\omega^2 a^2}=4 \dfrac{k_{\mathrm{phys.}}^2}{\omega^2}, \quad q= \mathcal{G}f^2/\omega^2.
\ee
A detailed analysis of the above Mathieu equation \cite{Mathieu-Eq} will appear in an upcoming paper \cite{Progress}, here we only present the main results. For $q\ll 1$ we have a the narrow band resonance all the time which yields production of relativistic $\chi$-particles. Since $k_{\mathrm{phys.}} = k/a$,  it is stretched  and the resonance condition $A_\k =1 \pm q$ for each mode is met only for a short period $\Delta t$
\be
2 H \Delta t=2\dfrac{\Delta k_{\mathrm{phys.}}}{k_{\mathrm{phys.}}} =  2 q.
\ee
In this short time interval  $n_{\p}$, the number of created $\chi$-particles of momentum $\p$, is given by
\be
\label{nk:const}
n_{\p}\equiv |\beta_\p|^2 \simeq \sinh^2(q^2 \alpha/4) = \sinh^2\big(\frac{\mathcal{G}^2 f^4}{4 H \omega^3}\big).
\ee
As we will see below the parameter space of  our model allowed by the CMB data requires $q^2\alpha\lesssim 1$ and $n_\k$ does not exhibit exponential enhancement.

One can compute the energy density carried by the $\chi$ particles (which are highly relativistic for $q\ll 1$):
\begin{align}
\label{chi-energy:1}
\rho_{\chi} &= \dfrac{1}{2\pi^2 a^3} \int_{0}^{k< a \omega/2}   dk \,k^2 \, \omega_\k \, n_\k\cr 
&=  \dfrac{\omega^4}{128 \pi^2} \, \sinh^2\big(\frac{\mathcal{G}^2 f^4}{4 H \omega^3}\big)\simeq \rho_{bg}\  \frac{(q \alpha)^4}{2^{11}\cdot 3\pi^2}\frac{\omega^2}{\mpl^2}.
\end{align}
That is, energy density transferred to highly relativistic $\chi$ particles (radiation)  reaches a constant value during inflation. Moreover, as we will see the range of parameters we choose to have a successful inflationary model implies $\rho_\chi\ll \rho_{bg}$. Therefore, the backreaction of resonant particle production on the background inflation trajectory is very small.

\vspace{-6mm}
\section{Resonant interactions and cosmic perturbations}\label{sec:4}
\vspace{-4mm}

We now explore the effect of resonant interactions on power spectrum and spectral tilt of scalar perturbations. To this end we perform the one-loop inflaton 2-pt function calculations in in-in formulation \cite{Progress}. In the unitary gauge $\delta \phi =0$, the interaction of inflaton with $\chi$ is only via the modulated mass term of $\chi$ field $m^2_{\chi}=\mu^2+\mathcal{G} f^2 \cos(\phi/f)$: 
\be
{\cal L}_{\mathrm{int.}}\simeq -\dfrac{1}{2} {\cal G} f^2 \cos (\phi/f) ~\chi^2.
\ee
During slow-roll regime the argument $\phi/f=\omega t$, where $\omega$ is almost a constant. To compute interactions we expand the inflaton field around the rolling background, $\phi=\omega f t+\varphi$, where $\varphi$ denotes the (subhorizon, quantum) perturbations of inflaton field.
The above interaction Hamiltonian may then be expanded as
\be\label{interaction-Hamiltonian}
{\cal H}_{int.} \simeq g_3(t) \vphi \chi^2+ g_4(t) \vphi^2\chi^2+\cdots,
\ee
where 
\be\label{couplings-g3-g4}
g_3(t)=-\frac12{\cal G} f  \sin \omega t,\qquad g_4(t)=-\frac14 {\cal G}  \cos \omega t.
\ee

Mode expansion of the $\chi$ field, recalling the production relativistic particles in the narrow resonance band, is
\be\label{chi-mode-expansion}
\chi_{\p}(\eta) \simeq \dfrac{-H \eta}{\sqrt{2p}} \left( e^{-i p \eta}+\beta_\p\ \theta(\dfrac{\alpha}{2}+p\eta) e^{+i p \eta} \right),
\ee
where $\eta=-e^{Ht}/H$ is the conformal time,  $p^2=\p\cdot \p=|\p|^2$, $\theta(\dfrac{\alpha}{2}+p\eta)$ is the condition for the mode with momentum $p$ to be excited by the time $\eta$ and $\beta_\p$ is given in \eqref{nk:const}. For the modes relevant to our loop computations below, $\omega_\p(\eta) = \sqrt{p^2+m^2 a^2}\simeq p$ and $\simeq$ in \eqref{chi-mode-expansion} refers to the fact that it is written for modes inside the horizon with $p\eta\gtrsim 1$. 
Properly normalized $\varphi$ mode function is
\be\label{varphi-mode}
\varphi_\k(\eta) = \dfrac{H}{\sqrt{2 k^3}} (i-k\eta) e^{-ik\eta}
\ee
at early times, where the modes are deep inside horizon, it simplifies to 
\be\label{DIH-varphi}
\varphi_\k(\eta) \rightarrow \dfrac{-H \eta }{\sqrt{2 k}} e^{-ik\eta}.
\ee
That is, we have used usual Bunch-Davies vacuum (initial) state for both of $\chi$ and $\varphi$ modes. Note also that
\be\label{propagator}
\text{Re}\ \varphi_{\k}(\eta)= \dfrac{H}{\sqrt{2 k^3}} (\sin k\eta-k\eta \cos k\eta)
\ee
is propagator of $\varphi$ on de Sitter space which vanishes for superhorizon modes $(k\eta\ll 1)$.

Let us first analyze the effect of cubic interactions \eqref{interaction-Hamiltonian} and the particles $\chi$ produced on the 2-pt function of inflaton fluctuations using the standard in-in formulation (see \cite{Progress} for details of computations). This appears as a one-loop effect given by the following integral
\be\label{2pf}
\hspace*{-2mm}\dfrac{\expect{\vphi_{\k} \vphi_{-\k}}_{\text{1-loop}}}{P_{\phi}(k)}= -4{\rm Re} \int d^3\p 
\int^0_{-\infty} \frac{d\eta_1}{H^4\eta_1^4} g_3 \text{Re}\vphi_{\k} \chi_{-\p}\chi_{\p-\k}\ {\mathbf{I}_0},\nonumber
\ee
where $P_{\phi}(k)= \expect{ \varphi_{\k} \varphi_{-\k}}_{\text{vac}}$ is the power of superhorizon vacuum fluctuations and
\be
{\mathbf{I}_0}={\mathbf{I}_0}(\eta_1,\k,\p)=\int^{\eta_1}_{-\infty}\frac{d\eta_2}{H^4\eta_2^4} g_3 \vphi^*_{-\k} \chi^*_{\p}\chi^*_{-\p+\k}.\nonumber
\ee
Recalling the mode expansion of $\chi$ \eqref{chi-mode-expansion} and that for the range of parameters we are interested in $\beta_\p \ll 1$, the dominant contribution to the 2-pt function comes from the ``one-loop vacuum'' (where there is no particle production) \footnote{Contribution of the quartic interaction term \eqref{interaction-Hamiltonian} to the one-loop 2-pt function can be shown to be negligible compared that of the cubic interactions we analyze here \cite{Progress}.
Moreover, the higher loop one-particle irreducible diagrams will still be subdominant because the resonance interactions effects discussed here do not take place. One may also wonder about the reducible higher loop diagrams. These are also suppressed as the propagator connecting each one-loop part, when each one-loop part is in resonance, is vanishing \cite{Progress}. Note that, each one-loop diagram involves two kinds of resonance effects, one on $\eta$ integrals and one on momentum integral. The details of these computations will apeear in \cite{Progress}.}
\footnote{We note that loop effects in inflation, including cases like ours which has modulations and resonances, has been studied in some interesting papers \cite{Loop-Inflation-1,Loop-Inflation-2,Loop-Inflation-3}.}. Our explicit computations of course confirms this \cite{Progress}. Therefore, 
\be\label{2pf-simp}
\dfrac{\expect{\vphi_{\k} \vphi_{-\k}}_{\text{1-loop}}}{ P_{\phi}(k)}= -\frac{{\cal G}^2 f^2}{8H^2}\ {\rm Re} \int \dfrac{d^3\p}{pk|-\p+\k|} \ {\mathbf{I}_1},
\ee
where 
\be\label{I-pk}\begin{split}
{\mathbf{I}_1} &=\int^0_{-\infty}  \frac{d\eta_1}{\eta_1} \sin \omega t_1 ~e^{-i(p+|\p-\k |)\eta_1}(\cos k\eta_1-\frac{\sin k\eta_1}{k\eta_1}) {\mathbf{I}_2},\cr
\hspace*{-2mm}{\mathbf{I}_2} &=\int^{\eta_1}_{-\infty} \frac{d\eta_2}{\eta_2} \sin \omega t_2 ~e^{i(k+p+|\p-\k |)\eta_2}(1+\frac{i}{k\eta_2}).\nonumber
\end{split}
\ee
The above integrals cannot be obtained in the closed form, while one can compute them in the stationary phase (saddle point) approximation:
\be\label{saddle-point:1}
\begin{split}
\hspace*{-3mm}\int^{\eta} \dfrac{d \eta'}{\eta'} \sin\omega t'\, e^{\pm i \ell \eta'}  
&\approx \mp\sqrt{\dfrac{\pi }{2\,\alpha}} e^{\pm i\Phi(\ell,\alpha)}\ \theta(\eta+\eta_s),\cr
\hspace*{-3mm}\int^{\eta} \dfrac{d \eta'}{\ell\eta'^2} \sin\omega t'\, e^{\pm i \ell \eta'}  
&\approx \pm\sqrt{\dfrac{\pi }{2\,\alpha^3}} e^{\pm i\Phi(\ell,\alpha)}\ \theta(\eta+\eta_s),
\end{split}
\ee
where
\be\label{Phi}
\begin{split}
\Phi(\ell, \alpha)&=\alpha \ln \alpha -\alpha-\alpha \ln (-\ell\eta_{\ast})+\frac{\pi}{4},\cr
\eta_s&\approx\eta_0(1+\frac{1}{\sqrt{\alpha}}),\qquad \eta_0=\frac{\alpha}{\ell},    
\end{split}
\ee
$\eta_{\ast}$ denotes some arbitrary time and $\theta(\eta+\eta_s)$ is the step function. 
The crucially important point in integrals $\mathbf{I}_1, \mathbf{I}_2$ is the appearance of oscillating time dependent couplings $\sin\omega t$ factors, leading to resonant one-loop effects. Computing the integrals in the steady phase approximation is then straightforward, but needs special care on considering the role of step functions, which results in 
\be
\text{Re}\ {\mathbf{I}_1}\simeq -\frac{\pi}{2\alpha} \left(\cos\frac{\alpha k}{\kappa_0}-\frac{\kappa_0}{\alpha k}\sin\frac{\alpha k}{\kappa_0}\right)^2
\ee
where $\kappa_0=p+|\p-\k|$ and we assumed $k\lesssim \kappa_0$ and $\alpha\gg 1$.
Finally we plug $\text{Re}\ {\mathbf{I}_1}$ into \eqref{2pf-simp} to obtain
\be\label{2pf-simp-fina}
\nonumber\begin{split}
\dfrac{\expect{\vphi_{\k}\vphi_{-\k}}_{\text{1-loop}}}{P_{\phi}(k)}&=\cr
 \frac{\pi\ {\cal G}^2 f^2}{16H^2\alpha}& \int_{p_{_{IR}}}  \dfrac{p^2dp\ d\Omega_p}{pk|-\p+\k|} \big(\cos\frac{\alpha k}{\kappa_0}-\frac{\kappa_0}{\alpha k}\sin\frac{\alpha k}{\kappa_0}\big)^2\cr
=\frac{\pi^2{\cal G}^2 f^2}{8H^2}&\int_A^\infty\ du \int_{\frac{2u-1}{\alpha}}^{\frac{2u+1}{\alpha}} \frac{dz}{z^2} (\cos z-\frac{1}{z}\sin z)^2\end{split}
\ee
where the IR cutoff $p_{_{IR}}=\text{max}(H, A {k})=Ak$ and $A\simeq 1+1/\sqrt{\alpha}$ is a number bigger than one. The integral is naturally cutoff at UV, as for large $p$, $\kappa_0\sim 2p$ and there is a oscil-log function. Note also that the main contribution comes from momenta $p\eta\sim \alpha$ for which $\omega_\p(\eta)=\sqrt{p^2+m^2a^2}\simeq p$, justifying \eqref{chi-mode-expansion}. Moreover, when
$\alpha k\ll \kappa_0\sim 2p$, $\mathbf{I}\propto \alpha^3/H^2\ k^4/p^4$, so the integral is UV finite. For large $\alpha$ the integral simplifies to,
\be\label{2pf-simp-final}
\begin{split}
\hspace*{-3mm}\dfrac{\expect{\vphi_{\k} \vphi_{-\k}}_{\text{1-loop}}}{ P_{\phi}(k)}&=\frac{\pi^2{\cal G}^2 f^2}{8H^2}\ \int_{\frac{2A}{\alpha}}^\infty  \frac{dz}{z^2} (\cos z-\frac{1}{z}\sin z)^2\cr 
&\simeq \frac{\pi^3}{96} \left(\frac{\mathcal{G}^2 f^2}{H^2}\right)\equiv \frac{\pi^3}{96} \left(\frac{\mathbb{H}}{H}\right)^2
\end{split}
\ee
The above is our main result and shows if $\mathbb{H}={\cal G} f\gtrsim H$, the one-loop result which is enhanced because of  resonance occurring due to oscillatory coupling, dominates over the usual tree level power spectrum.  Therefore, 
\be\label{P-R}
P_{\cal R}\simeq \frac{\pi}{768 \epsilon} \big(\frac{\mathbb{H}}{\mpl}\big)^2=\frac{\pi}{384}\frac{\mathcal{G}^2}{\alpha^2},
\ee
and for the spectral tilt we get
\be\label{n-s}
n_s-1=2\frac{d\ln{\cal G}}{d\ln\gamma}(\eta-\epsilon)+2\eta-4\epsilon.
\ee
In a similar way one can compute contributions of order $\beta_\k$ and $\beta_\k^2$ as well as the contribution from the $\varphi^2\chi^2$ interaction and show that they are subdominant when $n_\k \ll 1$. Since the created $\chi$ particle pairs carry negligible energy compared to the background inflaton, the analysis for tensor modes is not altered compared to the usual case, yielding
\be
P_T=\frac{2}{\pi^2} \left(\frac{H}{\mpl}\right)^2,\ \quad r=\frac{P_T}{P_{\cal R}}=16\epsilon\ \frac{96}{\pi^3} \left(\frac{H}{\mathbb{H}}\right)^2 
\ee
As we see $r$ is suppressed by $(H/\mathbb{H})^2$ factor compared to the Lyth bound $16\epsilon$. This, together with the $n_s$ \eqref{n-s} is what allows us to relax Planck bounds on the inflation models with convex potential.
\vspace{-6mm}
\section{Comparison with the data}
\vspace{-4mm}

\begin{figure}[ht]
\centerline{
\includegraphics[scale=0.6]{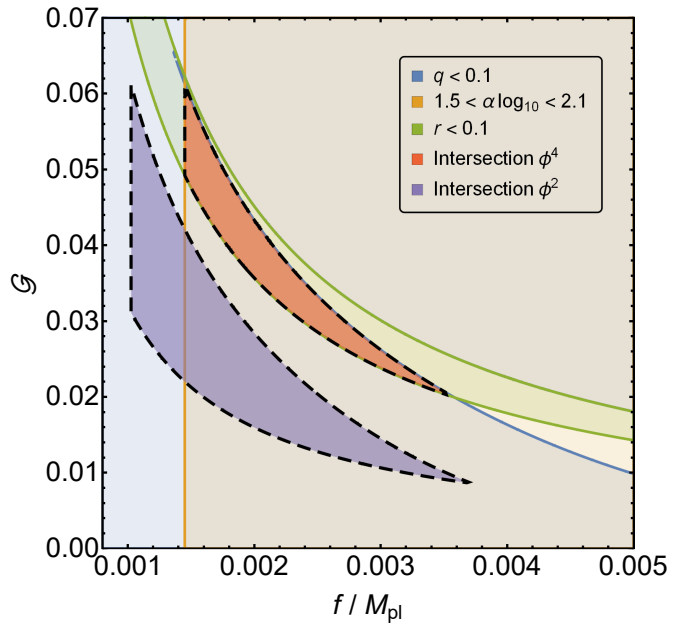}}
\caption{The allowed region of ${\cal G}$-$f/\mpl$ parameter space for $m^2\phi^2, \lambda\phi^4$ theories. The contour plots of $q, \alpha$ and $r$ are drawn for the $\phi^4$ case. For the $\phi^2$ we have only drawn the allowed region. In both $\phi^4, \phi^2$ cases the vertical bound on the left comes from the lower bound on $\alpha$, the lower bound curve  from $r<0.1$ and the upper bound curve from $q<0.1$.}
\label{Fig:g-f}
\end{figure}

There are four parameters $f/\mpl, \alpha, {\cal G}={\cal G}(\gamma), q$ in our model. The value for $P_{\cal R}$ essentially fixes 
${\cal G}/\alpha\simeq 5\times 10^{-4}$. Our model works and  our analysis is valid if $q\ll 1$. The $\cos(\phi/f)$ term in slow-roll parameter $\eta$, as discussed in \cite{omega-bound}, leads to a cosine-log term in spectral tilt $n_s$, yielding the bound $1.5<\log_{10}\alpha <2.1$. Requiring $r<0.1$, as implied by Planck data, restricts  ${\cal G}$ and $f/\mpl$ to a region roughly like $f/\mpl\sim 10^{-5}/{\cal G}$. The Planck data on spectral tilt $n_s = 0.9587 \pm 0.0056$,  for chaotic models with $\phi^4$ inflaton potential, assuming 60 e-folds of inflation, leads to $d\ln{\cal G}/d\ln\gamma=-1.1$-$-1.8$. From  Fig. \ref{Fig:g-f} besides ${\cal G}\sim 10^{-2}, f/\mpl\sim 10^{-3}$, one can read the range of Hubble parameter during inflation $H=0.5-2.5 \times 10^{-5}\ \mpl$, which is lower than the usual value for chaotic models ($3.7-5.2\times 10^{-5}\mpl$ for $p=2,4$ models).

\textbf{\textit{Discussion and outlook.}} We have presented a ``resonant module'' which could be added to arbitrary (slow-roll) models. In this way one can reduce $r$ and improve their behavior in comparison with the data. As most interesting examples we have shown how this works for the $\phi^2, \phi^4$ models. However, our scenario besides suppression of $r$ can also modify the spectral tilt \emph{cf.} \eqref{n-s} and hence can improve on other models already ruled out by the current data like the hybrid model \cite{hybrid}. We will show in a longer paper \cite{Progress} that: i) as \eqref{n-s} indicates, the first and second tilt of the tilt can be used to constrain derivative of ${\cal G}$ and, ii) while power spectrum is enhanced due to one-loop resonant effects, there is no such enhancement for non-Gaussianity for generic shape. Nonetheless, there could be enhancements and modulations for the specific case of folded non-Gaussianity, which may be compared with the analysis and results in \cite{Planck-2015-NG}. As we argued, the ratio of the energy carried by the resonant production of $\chi$-particles to background energy during inflation is very small. This ratio, however, grows as we come close to the end of slow-roll. The particle production of course continues after the end of inflation marked by $\epsilon=1$ and hence provides a natural setup for (p)reheating scenario \cite{preheating}. 

\textbf{\textit{Acknowledgements.}} We would like to especially thank Hossein Bazrafshan for his collaboration on early stages of this work and Mehrdad Mirbabaei for discussions and Eva Silverstein for comments on the draft. 
AAA acknowledges support of school of physics of IPM, where he is a part-time member. MMShJ was supported by the grants from ICTP NT-04, INSF grant No. 950124 and Saramadan grant No. ISEF/M/97219.  The authors acknowledge the associates office and HECAP sector of ICTP and the hospitality of their staff.



\begin{thebibliography}{99}

\bibitem{Planck-2018} 
  Y.~Akrami {\it et al.} [Planck Collaboration],
  ``Planck 2018 results. X. Constraints on inflation,''
  arXiv:1807.06211 [astro-ph.CO].

  
\bibitem{monodromy-SW}
E.~Silverstein and A.~Westphal,
  ``Monodromy in the CMB: Gravity Waves and String Inflation,''
  Phys.\ Rev.\ D {\bf 78}, 106003 (2008).
  [arXiv:0803.3085 [hep-th]].
 
 
 
  \bibitem{axion-monodromy-GW}
   L.~McAllister, E.~Silverstein and A.~Westphal,
  ``Gravity Waves and Linear Inflation from Axion Monodromy,''
  Phys.\ Rev.\ D {\bf 82}, 046003 (2010)
[arXiv:0808.0706 [hep-th]].



  \bibitem{axion-monodromy-CMB-features}
  R.~Flauger, L.~McAllister, E.~Pajer, A.~Westphal and G.~Xu,
  ``Oscillations in the CMB from Axion Monodromy Inflation,''
  JCAP {\bf 1006}, 009 (2010)
  [arXiv:0907.2916 [hep-th]].


\bibitem{Ngy-axion-monodromy} 
  S.~Hannestad, T.~Haugbolle, P.~R.~Jarnhus and M.~S.~Sloth,
  ``Non-Gaussianity from Axion Monodromy Inflation,''
  JCAP {\bf 1006}, 001 (2010)
  [arXiv:0912.3527 [hep-ph]].


\bibitem{monodromy-and-Planck} 
  R.~Easther and R.~Flauger,
  ``Planck Constraints on Monodromy Inflation,''
  JCAP {\bf 1402}, 037 (2014)
  [arXiv:1308.3736 [astro-ph.CO]].

\bibitem{monodromy-MSWW}
  L.~McAllister, E.~Silverstein, A.~Westphal and T.~Wrase,
  ``The Powers of Monodromy,''
  JHEP {\bf 1409}, 123 (2014)
  [arXiv:1405.3652 [hep-th]].

\bibitem{omega-bound} 
  C.~Zeng, E.~D.~Kovetz, X.~Chen, J.~B.~Muñoz and M.~Kamionkowski,
  ``Searching for Oscillations in the Primordial Power Spectrum with CMB and LSS Data,''
  arXiv:1812.05105 [astro-ph.CO].
  
  
\bibitem{Planck-2015-inflation}
  P.~A.~R.~Ade {\it et al.} [Planck Collaboration],
  ``Planck 2015 results. XX. Constraints on inflation,''
  Astron.\ Astrophys.\  {\bf 594} (2016) A20
  doi:10.1051/0004-6361/201525898
  [arXiv:1502.02114 [astro-ph.CO]].


 \bibitem{Planck-2015-NG}
  P.A.~R.~Ade {\it et al.} [Planck Collaboration],
  ``Planck 2015 results. XVII. Constraints on primordial non-Gaussianity,''
  Astron.\ Astrophys.\  {\bf 594}, A17 (2016)
  [arXiv:1502.01592 [astro-ph.CO]].

\bibitem{Mathieu-Eq}
N. McLachlan, \emph{``Theory and Applications of Mathieu 
Functions,''} (Oxford Univ. Press, Clarendon, 1947).

\bibitem{Progress}
A.A. Abolhasani, M.M. Sheikh-Jabbari, H. Bazrafshan Moghaddam, \textit{Work in Prrparation}.

\bibitem{Loop-Inflation-1} 

M.~S.~Sloth,
``On the one loop corrections to inflation and the CMB anisotropies,''
  Nucl.\ Phys.\ B {\bf 748}, 149 (2006)
  [astro-ph/0604488]; 
  ``On the one loop corrections to inflation. II. The Consistency relation,''
  Nucl.\ Phys.\ B {\bf 775}, 78 (2007)
  [hep-th/0612138].

 L.~Senatore and M.~Zaldarriaga,
``On Loops in Inflation,''
  JHEP {\bf 1012}, 008 (2010)
  [arXiv:0912.2734 [hep-th]];   ``On Loops in Inflation II: IR Effects in Single Clock Inflation,''
  JHEP {\bf 1301}, 109 (2013)
  [arXiv:1203.6354 [hep-th]].
  
G.~L.~Pimentel, L.~Senatore and M.~Zaldarriaga,
  ``On Loops in Inflation III: Time Independence of zeta in Single Clock Inflation,''
  JHEP {\bf 1207}, 166 (2012)
  [arXiv:1203.6651 [hep-th]].



\bibitem{Loop-Inflation-2} 
  
  P.~D.~Meerburg, M.~M\:unchmeyer and B.~Wandelt,
  ``Joint resonant CMB power spectrum and bispectrum estimation,''
  Phys.\ Rev.\ D {\bf 93}, no. 4, 043536 (2016)
  [arXiv:1510.01756 [astro-ph.CO]].

  X.~Chen, R.~Easther and E.~A.~Lim,
  ``Large Non-Gaussianities in Single Field Inflation,''
  JCAP {\bf 0706}, 023 (2007)
  [astro-ph/0611645].

  R.~Flauger and E.~Pajer,
  ``Resonant Non-Gaussianity,''
  JCAP {\bf 1101}, 017 (2011)
  [arXiv:1002.0833 [hep-th]].
 
  S.~R.~Behbahani and D.~Green,
  ``Collective Symmetry Breaking and Resonant Non-Gaussianity,''
  JCAP {\bf 1211}, 056 (2012)
  [arXiv:1207.2779 [hep-th]].
  
  S.~R.~Behbahani, A.~Dymarsky, M.~Mirbabayi and L.~Senatore,
  ``(Small) Resonant non-Gaussianities: Signatures of a Discrete Shift Symmetry in the Effective Field Theory of Inflation,''
  JCAP {\bf 1212}, 036 (2012)
  [arXiv:1111.3373 [hep-th]].
 

\bibitem{Loop-Inflation-3} 
  R.~Flauger, M.~Mirbabayi, L.~Senatore and E.~Silverstein,
  ``Productive Interactions: heavy particles and non-Gaussianity,''
  JCAP {\bf 1710}, no. 10, 058 (2017)
  [arXiv:1606.00513 [hep-th]].


\bibitem{hybrid} 
  A.~D.~Linde,
  ``Hybrid inflation,''
  Phys.\ Rev.\ D {\bf 49}, 748 (1994)
  [astro-ph/9307002].

\bibitem{preheating} 
  L.~Kofman, A.~D.~Linde and A.~A.~Starobinsky,
  ``Towards the theory of reheating after inflation,''
  Phys.\ Rev.\ D {\bf 56}, 3258 (1997)
  [hep-ph/9704452].

 P.~B.~Greene, L.~Kofman, A.~D.~Linde and A.~A.~Starobinsky,
  ``Structure of resonance in preheating after inflation,''
  Phys.\ Rev.\ D {\bf 56}, 6175 (1997)
  [hep-ph/9705347].

\end{thebibliography}
\end{document}